\documentclass[twocolumn,showpacs,preprintnumbers,amsmath,amssymb]{revtex4}
\usepackage{graphicx}% Include figure files
\usepackage{dcolumn}% Align table columns on decimal point
\usepackage{bm}% bold math
\begin{document}
%\draft

\title{ Impurity-Semiconductor Band Hybridization Effects on the Critical Temperature of Diluted Magnetic Semiconductors.}

\author{M. J. Calder\'on$^{1,3}$, G. G\'omez-Santos$^2$, and L. Brey$^3$. }
\affiliation{\centerline{$^1$Cavendish~Laboratory, Cambridge~University,~Madingley~Road,~Cambridge CB3 0HE, UK. }}
\affiliation{\centerline {$^2$Departamento~de~F\'{\i}sica~de~la~Materia~
Condensada and Instituto~Nicol\'as~Cabrera,~Universidad~Aut\'onoma~de~Madrid,~28049~Madrid,~Spain.}}
\affiliation{\centerline{$^3$Instituto de Ciencia de Materiales de Madrid (CSIC),~Cantoblanco,~28049,~Madrid,~Spain.}}

%\maketitle
\begin{abstract}

We have studied the critical
temperature of Diluted Magnetic Semiconductors by means of Monte Carlo
simulations and Coherent-Potential-Approximation (CPA) calculations. In our model for this system, the magnetic 
ions couple with the carriers through an antiferromagnetic exchange interaction, $J$, and
an electrostatic interaction $W$. The effective impurity potential $J-W$ controls
the hybridization between the magnetic impurities and the hole charge on the dopants.
We find that the critical temperature depends substantially on the hole charge 
on the magnetic impurities. 
The CPA critical temperature is always lower than the obtained in the Monte 
Carlo simulations, although all trends in the simulation results are reproduced
in the CPA calculations. Finally we predict the existence of pockets of phase segregation
instability close to the carriers band edges.

\end{abstract}

\maketitle

\section {Introduction.}
It has been recently discovered  that some Mn doped semiconductors, such  as
Ga$_{1-x}$Mn$_x$As,
undergo a ferromagnetic-paramagnetic transition at temperatures near 100K \cite{Matsukura,Matsukura-bis}.
Motivated by this high Curie temperature, there have been further experimental 
studies in III-V magnetic semiconductors, and 
ferro-paramagnetic transition temperatures as high as 1000K have been obtained in (Ga,Mn)N \cite{Sonoda}.
 The optimal Mn concentration is near 5$\%$ in these materials, making  magnetic
ions to be  rather diluted in the host semiconductor.

The existence of ferromagnetic semiconductors at room temperature opens the 
possibility of combining the spatial modulations of both carrier  and  
spin density.
Therefore, diluted magnetic semiconductors (DMS) have attracted much attention for their 
potential use in spintronic devices.
Some spintronic devices have been already 
implemented \cite{Koshihara},
and different heterostructures in order to optimize and control the critical temperature have
been proposed \cite{Loureiro,Rossier,Brey,Jungwirth}.

In this work we study magnetic properties   III-V magnetic semiconductors, 
in particular
of GaMnAs. 
The high critical temperature DMS's have a high concentration
of randomly located manganese ions. From electron paramagnetic resonance
and optical experiments \cite{Szcytko,Linnarson}, it is known that the Mn ions in 
these  semiconductors have a $S=5/2$ local moment. 
The Mn ions go in substitutionally at the cation sites, Ga, and contribute holes
to the semiconductor valence band. 
Although the local moment of the magnetic ions
in GaMnAs indicates that each Mn should contribute with a hole,  it is experimentally 
observed that the density of itinerant holes, $p$, 
is a small fraction ($\sim 10 \%$) of the Mn concentration \cite{Ohno}.
The reduction in the carrier density  is probably due to the presence of antisite defects
which, acting as deep donors partially compensating the Mn acceptors, 
reduce the hole density \cite{Ohno1}.
A large number of antisite defects is expected to be present in these systems
 because, in order to minimize the mobility of
the Mn ions and the existence of  clusters, the semiconductors are
fabricated  using molecular beam epitaxy growing techniques at low temperatures,
for which a large number of these defects is known to occur \cite{Ohno1,VanEsch}.

It is widely accepted that the ferromagnetism in III-V semiconductors is induced by
the motion of the electrical carriers in the system. However, there is no agreement
on the precise role  of the carriers. 
They can be  either itinerant, with a clear character of the  host semiconductor
valence band states \cite{Dietl,Dietl-bis,Konig,Jungwirth1},
or rather they can be mostly located on Mn impurities, with a clear
impurity band magnetic, polaron character \cite{Berciu,Berciu-bis,Kaminski,Yang}.
On the other hand, the carrier  induced ferromagnetic interaction 
between Mn$^{2+}$ ions in 
magnetic semiconductors has some resemblance to the one observed in
manganese perovskites \cite{Tokura}, double perovskites \cite{Kobayashi}, 
and pyrochlores \cite{Shimikawa}.
In these systems, the  phase diagram is different from that of conventional
itinerant ferromagnets, showing first-order transitions, 
phase separation \cite{Guinea}, and formation of
magnetic polarons near the Curie temperature \cite{Majumdar,Majumdar-bis}.

   In order to study these issues, we use a tight-binding model Hamiltonian 
in which, besides the Hund's coupling, a site energy 
difference between magnetic and normal ions, $W\!$, is introduced. Monte Carlo (MC)
  simulations are then performed for
this model, with particular emphasis on the $W$ dependence of the phase diagram.
 A mean field calculation within the 
coherent-potential-approximation (CPA)
is then carried out and compared with MC results. The CPA treatment confirms the
 robustness of
the MC results for the effect of $W\!$, in addition to allowing a simpler exploration 
of important issues such as the existence of phase separation.

Our main results are the following:
\par \noindent 1) The hole charge on the Mn ions, controlled by the impurity potential $W$,
 has a profound effect on the transition temperature. Upon increasing $W$, the hole charge on
 Mn decreases, leading to a substantial decrease of the Curie temperature.
\par \noindent 2) For much of the model's parameter space, the MC critical 
temperature is higher than that obtained using the simplest Weiss mean field theory.
The MC critical temperature is {\it always} higher than the corresponding CPA temperature.
Nevertheless, all trends in the MC results, particularly
the effect of the impurity potential, are reproduced in the CPA calculation. We believe this to
be a robust feature.
 \par \noindent 3) The CPA calculation shows the existence of small pockets
  of phase segregation instability close to band edges.  

The paper is organized in the following way. In section II we introduce and explain the
model.
Section III and IV  describe the MC and  CPA 
methods, respectively. In Section V we present our results, both MC and CPA,
making contact with
previous calculations and/or experiments.  Section VI summarizes our work.

\section {The model.}

It is generally accepted \cite{Dietl,Dietl-bis,Konig,Jungwirth1,Berciu,Berciu-bis,Kaminski}
that the DMS's are governed by the following Hamiltonian:
\begin{equation}
H=H_{h}+H_{ions}+H_{h-ions} \, \, \, \, \, \, ,
\label{Htotal}
\end{equation}
$H_h$ and $H_{ions}$ are the   parts of the Hamiltonian  describing  the 
holes and the Mn ions, respectively. 
The term $H_{h-ions}$ represents the interaction between the magnetic ions and the carriers.

$H_h$  is the sum of the kinetic energy of
the carriers, 
the hole-hole interaction energy, and the interaction of the holes
with the disorder potential from randomness in the host semiconductor, 
in particular from the presence of antisite defects.
In the actual magnetic semiconductors, the carrier density  is of the order
of $10^{20}$ cm$^{-3}\!$, and, for these values, it is justified to neglect the effect
of the carrier-carrier interaction. 
In this work we do no take into account the interaction of  
carriers with antisite disorder. This contribution is probably   important  in 
actual samples, but here  we are mainly interested in the effect  of the
on-site energy difference and  thermal fluctuations on  the
critical temperature.
We describe the motion of the holes using a single one-band hole dispersion, instead of the
more appropriated six bands envelope function formalism \cite{Albolfath}. This one-band 
model  is probably insufficient to give a quantitative prediction for the Curie
temperature, but  it includes the main features of the hole system we want
to study:  the effect of thermal fluctuations, Mn-GaAs orbital hybridization,  
and charge localization effects.
With all this, the Hamiltonian for the holes is written as
\begin{equation}
H_h = - t \sum _{< i,j >} \left ( C ^+ _{i,\sigma} \,  
C  _{j,\sigma} + h.c. \right ) \, \, \,,
\label{hamilholes}
\end{equation}
where the sum runs over all first neighbors pairs on a simple cubic lattice, $t$ is the
tunneling amplitude, and $ C ^+ _{i,\sigma}$ creates a hole with spin $\sigma$ at site $i$.
This model has a hole bandwidth of $12t$, and the parameter $t$ has been chosen  to give  
a bandwidth of the order of that of GaAs ($\sim $10$eV$), implying $t \sim$0.84$eV$.
The band edges of this model correspond to a parabolic dispersion with an effective mass
$m ^* = \hbar ^2 / 2 a ^2 t$, being $a$ the lattice parameter of the simple cubic lattice.
The value of $a$ has been  chosen to make  $a^3$  the volume per GaAs unit, i.e. 
$a$=3.56\AA. 
With these parameters, we obtain an effective mass $m ^*$=0.36$m_0$.

The term $H_{ions}$ in Eq.(\ref{Htotal}) describes the direct antiferromagnetic interactions 
between the magnetic moments of the Mn ions. Those are much smaller than the interactions
with the carrier spins and, therefore, we {\it neglect} them.

Finally, the term $H_{h-ions}$ represents the coupling between the electrical carriers 
and the magnetic impurities. There are two contributions to this term: i) An antiferromagnetic
exchange interaction between the spins ${\bf S} _I$ of the Mn$^{2+}$ ions located
at sites $I$ and the spins of the itinerant carriers. 
This interaction produces a long range ferromagnetic interaction between the Mn ions.
And ii) an interaction
between the carriers charge and the potential arising from the Mn dopants:
\begin{equation}
H_{h-ions}= { \frac J  S} \sum _{I} {\bf S} _ I  \, C ^ + _{I, \alpha} \, 
{\mbox {\boldmath $\sigma$}} _{\alpha, \beta} C _{I , \beta} \, +
\,  
W \sum  _{I}  C ^ + _{I, \sigma} \,  C  _{I, \sigma} \, \, \, .
\label{hamil_ion_h}
\end{equation}
Here the sum runs over the positions of the magnetic impurities, and 
${\mbox {\boldmath $\sigma$}}$ are the Pauli matrices. The origin of $W$ is Coulombic, coming from the 
different electronegativity of Mn and GaAs atoms, and from the screening of the impurities
by the electrical carriers.The quantity $J-W$ governs the
 hybridization of the magnetic impurities with the host bands, and the excess
of charge on the impurities with respect the host semiconductor  atoms. For instance,  
in the case of a perfect ferromagnetic ground state, the electrical charge is uniformly
distributed for $J$=$W$.
The value of $J$ in our model is connected with the exchange  coupling $J_{pd}$ appearing 
in the continuum model \cite{konig1}
through $J=SJ_{pd}/2a^3$. The value of $J_{pd}$ 
is typically 0.05-0.1$eV \, nm^3 $ \cite{Matsukura,Matsukura-bis,Omiya,Okabayashi}, 
yielding $J \sim$1.4-2.8$eV$, which gives a value of $J/t \sim$1.7-3.4.
Local spin density calculations (LSDA)  and supercell methods \cite{Sanvito} applied to 
GaMnAs predict a value of $J$ rather larger than the experimental one,
partially  because the LSDA calculations overestimate exchange coupling
strengths.
Lacking reliable experimental information on the value of $W\!$, 
 we consider it  as a parameter 
with values between $0$ and $J$.
Summarizing the model: we restrict ourselves to
the terms $H_{h}$ and $H_{h-ions}$, given by Eq.~\ref{hamilholes} and 
Eq.~\ref{hamil_ion_h}, respectively.

\section{Description of the Monte Carlo Algorithm.}
 We have performed classical MC simulations on the Mn core spin angles in order
to calculate the temperature, $T$, dependent magnetic phase diagram. The simulations
are done in a $N \times N \times N$ cubic lattice with periodic boundary conditions, and
 $N_{Mn}$=$xN^3$ randomly located
magnetic ions. The number of carriers is taken to be $N_h$=$pN ^3 $ . 

The simulations  are  done as follows. First, we produce a realization of the 
disorder:  the positions of the Mn ions. For this realization,  we compute the expectation value of the hole spin orientation at every place,
\begin{equation}
{\bf n  } _ i = < C ^+ _{i, \alpha } {\mbox {\boldmath $\sigma$}} _{\alpha, \beta} C _{i , \beta} >
\, \, \, .
\label{spin density}
\end{equation}
At a given temperature, the Mn spins fluctuate thermically on a fictitious magnetic field \cite{nota1}
proportional to the carriers spin density, 
\begin{equation}
{\bf B } _f (I) = {\frac  J S}  {\bf n } _ I \,  \, \, \, .
\end{equation}
We perform MC simulations on the Mn spin angles until the system is equilibrated, when  
we choose a particular core spin configuration, and recalculate the hole spin density.
 This cycle is repeated  until convergence in physical quantities is attained.
Calculating the hole spin density requires the diagonalization of the Hamiltonian
Eq.(\ref{Htotal}) at each step of the  cycle. ${\bf n } _ i$ is evaluated by filling up
the $N_h$ lowest energy levels of the Hamiltonian. This can be done because the Fermi temperature
is much higher than any other temperature in the system. For instance, for $x$=10$\%$,
$p$=1$\%$, and $J$=$W=$3$t$, the Fermi energy is near 0.5$t$, while the Curie temperature
is $\sim$0.015$t$.

The CPU diagonalization  cost grows as $N^6 $, severely limiting the system's sizes under study. 
By diagonalizing the Hamiltonian in real space, it is not possible to analyze
systems bigger than $N$=6 \cite{MJC,JAV}.
In order to study bigger systems, it is better to use an adapted basis rather than the 
atomic localized basis. In the simulation, we use the basis obtained by diagonalizing the hole
Hamiltonian for a perfect  ferromagnetic alignment of the Mn ion spins.
For the updating of the electron spin density at each cycle of the simulation,
it is not necessary to include all the eigenvectors of the extended basis.
We have found enough to work
with a reduced basis formed by the $N_c$ lowest energy eigenstates of 
this basis for each carrier spin orientation. 
The value of $N_c$ depends on the particular values of the parameters. 
In general, we find that 
the size of the required basis increases with the value of $J$-$W$, i.e. it increases with the
degree of localization of the hole charge on the Mn ions.
Typically we can obtain good convergence with relatively small values of $N_c$, most of
the results presented in this paper have been obtained with $N_c$=57.
Note that, in the extended basis we use, 
the square of the wave functions is not spatially uniform, as it would happen in a Bloch or plane wave basis,
but it is modulated by 
the spatial fluctuations  of the potential felt by the carriers. 
Using this extended basis and an appropriate cutoff $N_c$, we can perform MC simulations in systems with sizes up to N=16.

In the simulations, we calculate the thermal averages of different quantities:  
the absolute value of the Mn ions spin polarization,
\begin{equation}
M = {\frac 1 { S N_{Mn}}} \langle \, \, | \sum _I {\bf S } _ I |\, \,  \rangle \, \, \, \, ,
\label{Mn_magn}
\end{equation}
the absolute value of the hole's spin polarization, 
\begin{equation}
m = {\frac 1  {N_{h}}} \langle\, \,  | \sum _i {\bf n } _ i | \, \, \rangle \, \, \, \, ,
\label{hole_magn}
\end{equation}
and the average value of the hole charge on the Mn ions,
\begin{equation}
Q_{Mn} = {\frac 1  {N_{Mn}}} \langle  \, \,   \sum _I \rho  _ I \, \,  \rangle \, \, \, \, ,
\label{carga_mn}
\end{equation}
being 
$\rho _i = < \sum _{\sigma}  C^+ _ {i,\sigma} C _{i, \sigma} > $ the electrical 
 hole charge on site $i$. 
In the previous expressions, the index $I$ runs over the Mn positions and $i$
over all lattices sites.

\section{CPA method.}
\label{CPA}

  We have performed mean field calculations for  the present model within the
  framework of the Coherent-Potential-Approximation (CPA) \cite{t67,s67,ekl74}. 
  In recent times, this formalism has
  come to be known  as the
  Dynamical-Mean-Field-Approximation (DMFA) \cite{gkkr96}, and
  we use both terms (CPA and DMFA) as synonymous expressions. Given the possible confusion 
  from the use of 
  different flavors and names under the common umbrella of mean field 
  calculations, we devote this section to briefly describe our CPA formalism,
  pointing out the origin and nature of the approximations and, where appropriate, making
  contact with other approaches. 
  
 Tracing out the fermions at constant chemical potential and temperature,
  we end up with a
  probability distribution for the core spins given by the following 
  Boltzmann weight: 
\begin{equation}
\mathcal{P}(\left\{ \mathbf{S}_{i}\right\} )\propto exp\left[ -\beta \: \Omega
(\{\mathbf{S}_{i}\})\right]
\end{equation}
where $ \Omega (\{\mathbf{S}_{i}\}) $ is the carriers grand canonical potential
 for a given core-spin configuration.
 The {\it exact} distribution $ \mathcal{P}(\{\mathbf{S}_{i}\}) $ minimizes the
 following thermodynamical potential,
 considered as a functional over spin distributions:
\begin{equation}
\label{functional}
\mathcal{F}\left[ \mathcal{P}\right] =
\left\langle \Omega (\{\mathbf{S}_{i}\})\right\rangle +
T\left\langle \log (\mathcal{P})\right\rangle 
\end{equation}
where averages are taken over  core-spin distributions. This formulation 
offers a convenient starting point for approximate variational approaches. 
  
  The common basis  of all mean field  calculations is the
  minimization of Eq.~\ref{functional} with  a factorization {\it ansatz} for
   the spin distribution:
\begin{equation}
\mathcal{P}(\{\mathbf{S}_{i}\})=\prod _{i}p\left( {\bf S}_{i}\right)
\end{equation}

  From the critical phenomena point of view, this factorization  legitimizes the
  term mean field, implying that all criticality (if any) will be classical.
  Carrying out the minimization of Eq.~\ref{functional} with the only simplification of
  the factorization could  be termed the {\it absolute} mean field 
  approximation.
   Unfortunately, this procedure leaves us with the task of calculating
  the grand canonical potential for fermions ({\it i.e.} the density of states) in a
  site-disordered background: a formidable problem
  requiring  numerical treatments (see, for instance, Ref.~\cite{afglm01} for such a treatment
  in the Double-Exchange model). 
  
  To cope with the previous difficulty, additional approximations  are
  needed to obtain the average density of states for the disordered problem. We
  will describe two of them, namely, the Virtual-Crystal-Approximation 
  and  CPA treatment (see Ref.~\cite{ekl74} for a review on disordered systems).

  i) Virtual Crystal Approximation (VCA). The average density of states for
   the  disordered system is replaced by that 
    of the average Hamiltonian. That implies a translationally invariant system
    with site energies given by the  following average (a $2 \times 2$ matrix in spin space):
%\begin{equation}
$\varepsilon _{\alpha ,\beta }=
x (W+J\frac{\left\langle \mathbf{S}\right\rangle }{S}\cdot 
\mbox{\boldmath $ \sigma$} _{\alpha ,\beta })
%\end{equation}
$.
 The minimization of $ \mathcal  F$ is now straightforward, leading to the
   following well-known expression for the transition temperature: 
\begin{equation}
\label{TcVCA}
T_c  ^{VCA}= \frac{2}{3} x J^{2} n_{\sigma}(\mu)
\end{equation}
where $ n_{\sigma}(\mu) $  is the density of states per site and spin at the Fermi level 
(degenerate fermions assumed).
This scheme is often referred to as the Weiss mean field  approach, and we will
consider both expressions (Weiss mean field and VCA) as synonymous terms.
 
 ii) Coherent Potential Approximation (CPA). The CPA is a time-honored approach for
  the site-disordered "alloy" problem \cite{t67,s67}. It is  known to produce good results,
 particularly in three dimensions \cite{ekl74}. The averaged Green's function is obtained from 
 a translationally-invariant,  effective
 medium,  characterized by  a (site-diagonal and 
 energy dependent) self-energy $ \Sigma_{\alpha,\beta}(z)$ in the following
 manner:
\begin{equation}
\left\langle G(z)\right\rangle \simeq G^{^{CPA}}=(z-H_{_{_{CPA}}})^{-1}
\end{equation}
 where (Dirac notation used to emphasize the one-body character of the problem)
\begin{equation}
H_{_{CPA}}=
-t\sum _{( i,j)}
\left| i,\sigma \right\rangle \left\langle j,\sigma \right| 
+\sum _{i}\left| i,\alpha \right\rangle \Sigma _{\alpha ,\beta }(z)
\left\langle i,\beta \right|
\end{equation}
with sum over spin indices implied.
Given a particular  site-disorder distribution,  the self-energy is obtained 
 with a {\it minimum scattering} requirement: the substitution of an
 effective site by a real one should produce zero (local) scattering on the 
 average \cite{e83}. That is:
\begin{equation}
\left\langle (\hat{\varepsilon} _{i}-\hat{\Sigma} )
[\hat{1}-\hat{G}_{ii}^{^{CPA}}(\hat{\varepsilon} _{i}-\hat{\Sigma} )]^{-1}
\right\rangle =0
\end{equation}
where  {\it hats}  mean  $2 \times 2$ matrices in spin space, and the random quantities
$ \hat{\varepsilon_i} $ are given by:
\begin{equation}
\varepsilon _{i(\alpha, \,\beta )}=\left\{ \begin{array}{cc}
0, & probability=(1-x)\\
W + J\frac{\mathbf{S}_{i}}{S}\cdot 
\mbox{\boldmath $ \sigma$}_{\alpha, \,\beta }, & probability=x\: p\left( \mathbf{S}_{i}\right) 
\end{array}\right.
\end{equation}
 
%\end{document}  
  
    As usual, the minimization of the  thermodynamic potential can be cast
 in the form of a self-consistency equation for the probability distribution:
 \begin{equation}
\label{dmf1}
p\left( \mathbf{S}_{i}\right) \propto 
exp\left[ -\beta \: \delta \Omega (\mathbf{S}_{i})\right]
\end{equation}
where  $\delta \Omega (\mathbf{S}_{i})$ can be interpreted as the total change in the 
fermion's grand canonical potential for embedding an impurity site in the effective medium:
 \begin{equation}
\label{dmf2}
\delta \Omega (\mathbf{S}_{i})=\int d\omega \: f(\omega )
\frac{1}{\: \pi }\Im 
\log \det \hat{M}({\bf S}_i) 
\end{equation}
$f(\omega) $ is the Fermi-Dirac factor, $\Im$ means imaginary part, and $ \hat{M}({\bf S}_i)$ is 
given by:
\begin{equation}
 \hat{M}({\bf S}_i) =
 \hat{1} - \hat{G}^{^{CPA}}_{ii}(\omega ^{+}) \: [\hat{\varepsilon} (\mathbf{S}_{i})-\hat{\Sigma} (\omega ^{+})]
\end{equation}
  It is through eqs. \ref{dmf1} and \ref{dmf2}  that the CPA
  can be connected with the DMFA (see, for instance, Ref.~\cite{ak01}), an approach 
  primarily intended for  
  genuine many-body problems like the Hubbard model \cite{gkkr96}.
  It has been recently applied to models of itinerant carriers with ion
  impurities, believed to be relevant for manganites and doped magnetic
  semiconductors (see, for instance, Ref.~\cite{f95-1,f95-2,ak01,Chatto}). 
 
 Before leaving this section, let us comment on some features of these
 approaches. It is clear that the VCA is simple and handy, but the approximation in
 the density of states is too drastic. For instance, it predicts a ferromagnetic
 transition for any finite density of states.
  This cannot be true, even at the mean field level: for $W=0$ and
 small $J$, the problem becomes the standard
 Ruderman-Kittel-Kasuya-Yosida
 (RKKY) Hamiltonian,
 whose  {\it absolute} mean field treatment shows ferromagnetism only 
 for carrier density (holes or electrons) below a critical value  
 $p_c=0.5038$ (sc lattice), due to the oscillatory behavior of the spin-spin coupling.
 Furthermore, the VCA is not sensitive to the value of $W\!$, obviously a major limitation
 for this problem.
 In contrast, the CPA  is free from these shortcomings. For instance, it is  able to
 reproduce the {\it absolute} mean field results 
 of the RKKY limit \cite{ak01rkky}. In addition, extensive numerical results  available for 
 the exact  density of states 
 in the closely related Double-Exchange problem \cite{afglm01}, show virtually no 
  differences with the CPA density of states \cite{g02}. This  reinforces our 
  confidence in this method for the present problem.

\section{Results.}
\subsection{Monte Carlo Results.}
\begin{figure}
\includegraphics [clip,width=8.cm]{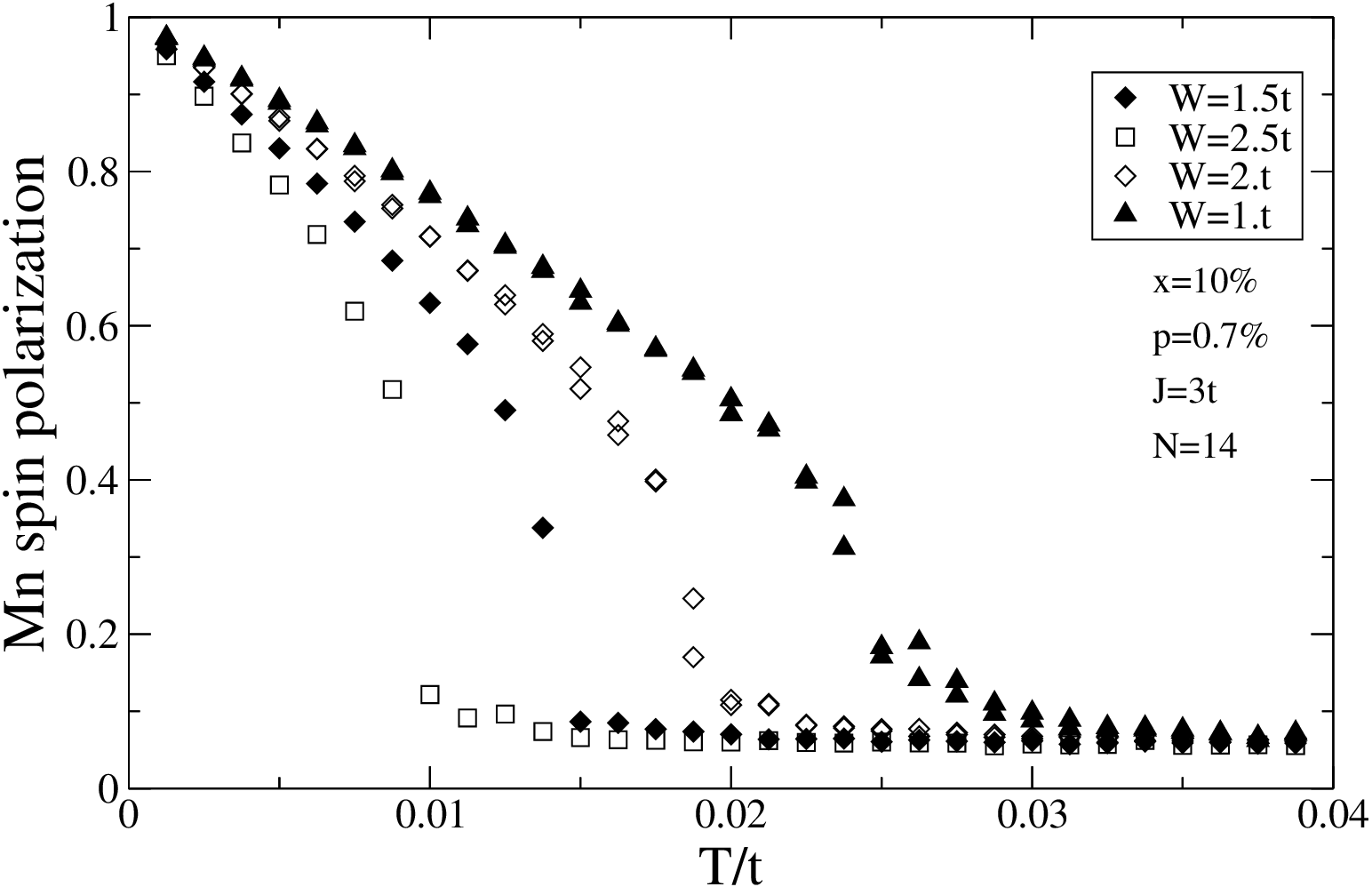}
\caption{Mn ions spin polarization as a function of temperature for 
$J$=3$t$, $x$=10$\%$, $p$=0.7$\%$, $N$=14 and  different values of the 
parameter $W$. Results for different realizations of disorder are plotted.} 
\label{fig1}
\end{figure}

\begin{figure}
\includegraphics [clip,width=8.cm]{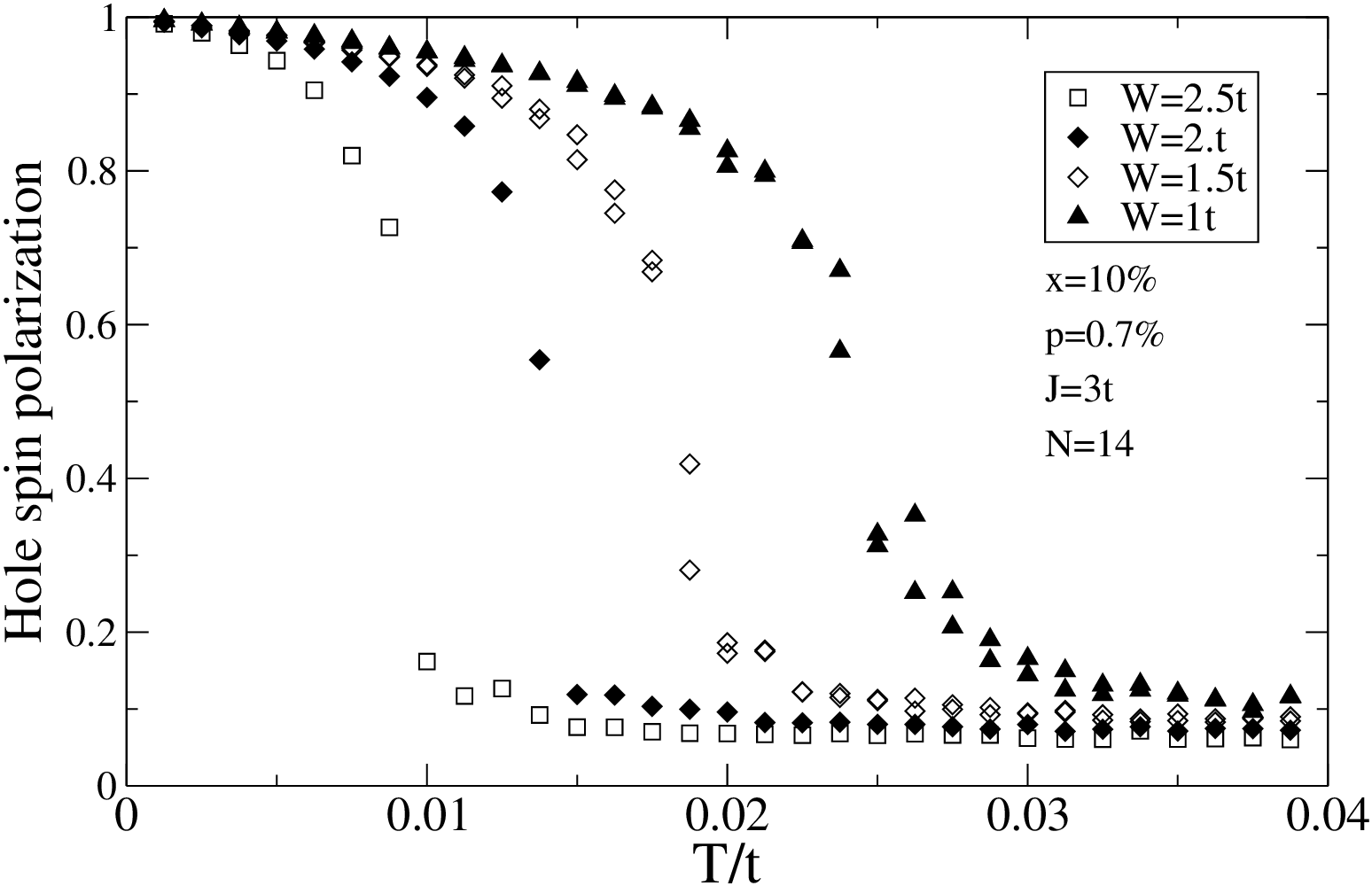}
\caption{Holes  spin polarization as a function of temperature for 
$J$=3$t$, $x$=10$\%$, $p$=0.7$\%$, $N$=14 and  different values of the 
parameter $W$.
Results for different realizations of disorder are plotted.
}
\label{fig2} 
\end{figure}

In Fig.~\ref{fig1} we plot the Mn ion's spin polarization, $M$, as a function of $T$,  for  $J$=3$t$, and 
different values of $W$. The results correspond to a Mn concentration $x$=10$\%$, a hole
concentration $p$=$0.7\%$, and a unit cell size  $N$=14. In Fig.~\ref{fig2} we plot the hole's spin
polarization for the same set of parameters.
Due to the finite size effects (note that there are only 274 Mn ions)
the polarizations $M$ and $m$ are different from zero at any temperature, and
we define the critical temperature, $T_c$, as the point where the second derivate of the polarization
with respect to $T$ changes sign. This way of obtaining $T_c$ implies uncertainties of around 5$\%$. 
Within this error margin, the $T_c$'s obtained from the temperature dependence of $m$ and $M$  coincide.
This is the expected result as the carrier's spin polarization is due to the Mn polarization.

Comparing Fig.~\ref{fig1} and Fig.~\ref{fig2}, we see that the ion spin polarization
decreases faster with temperature than that of the holes.
This seems to be generally true for the 
range of hole and Mn densities of interest, and a value of $J$=3$t$. 
This behavior has been already found in Weiss mean field calculations \cite{Jungwirth}.
Different disorder realizations are superimposed in figures 1 and 2.
 We find that, for small values of $J$-$W$, different disorder realizations 
give practically the same polarization curves, whereas, for small values
of $W\!$, fluctuations related with disorder become more important. This is because the larger
 $J$-$W$, the deeper
the effective impurity level. This increases the charge localized in the Mn ions, making
 the fluctuations associated with impurity randomness more important.

\begin{figure}
\includegraphics [clip,width=8.cm]{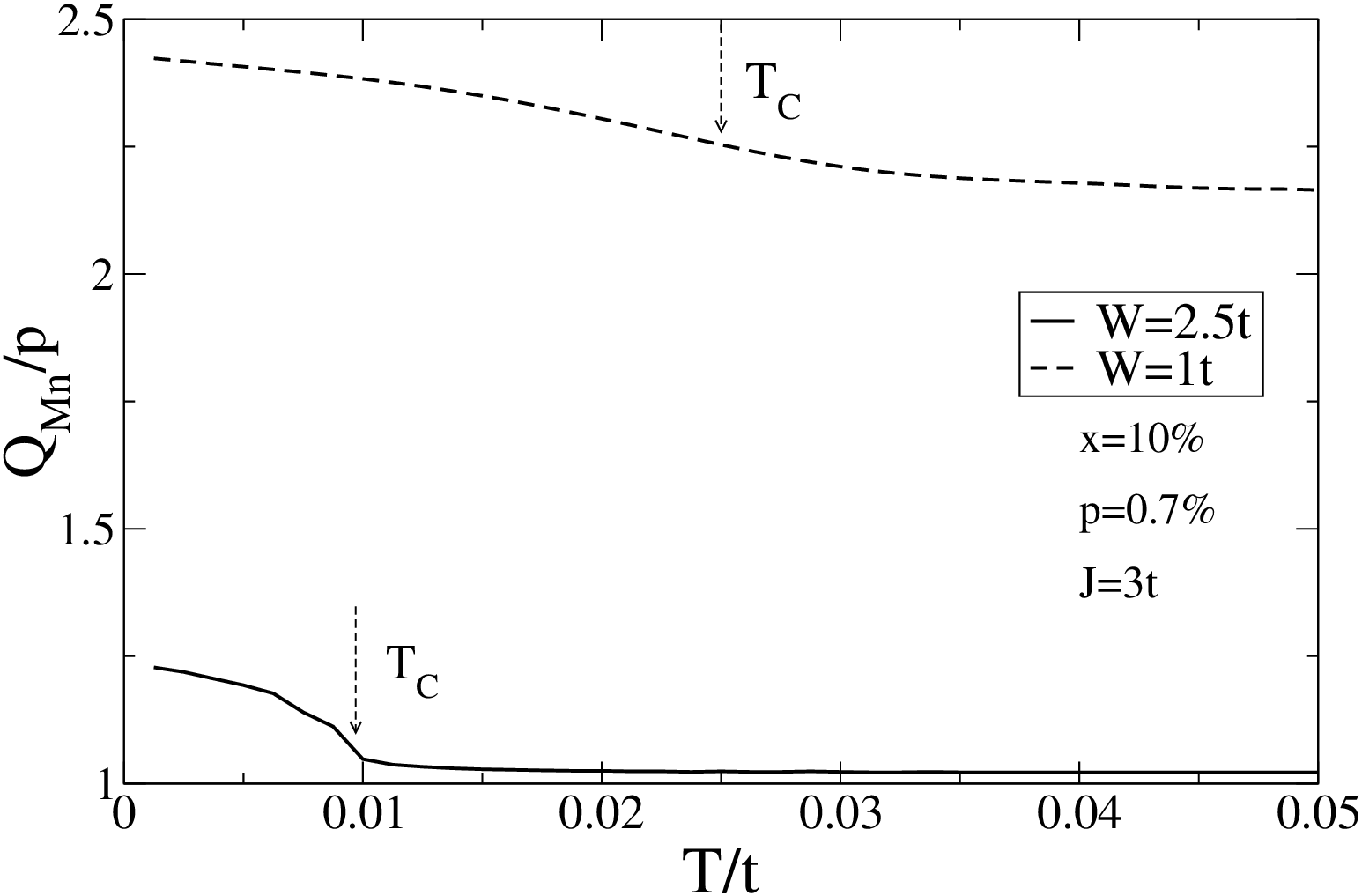}
\caption{Average hole charge on Mn ions, as a function of $T$, for two
different values of $W$. The results corresponds to the 
case $J$=3$t$, $x$=10$\%$, $p$=0.7$\%$, and $N$=14.
The arrows mark the Curie temperature, as obtained from Fig.~\ref{fig1} and Fig.~\ref{fig2}.
} 
\label{fig3}
\end{figure}

The critical temperature decreases when $W$ increases. As we will show,  this is
related to 
the amount of hole charge localized on Mn ions.
In Fig.~\ref{fig3} we plot the average hole charge on Mn  as a function of $T$, for two different values
of $W$.  Fig.~\ref{fig3} shows that  the charge on the Mn's decreases as the effective impurity
potential increases, that  $Q_{Mn}$  decreases with temperature, 
and that there is a change in the sign of the second derivate of $Q_{Mn}$ with respect to
temperature at $T_c$.

The delocalization of  charge with increasing $T$, and its abrupt change at $T_c$, 
indicate a possible change in  transport properties
of the system near the ferro-paramagnetic transition:
we expect that the electrical resistivity of the system will decrease when $T$ increases.
For a value of $W$=$J$, the effective impurity potential is exactly zero in the perfect
ferromagnetic state. Close to this value,   the charge on each Mn ion
is practically $p$, independently of $T$. However, for values of $W$ near zero, the charge on 
the Mn ions is very large, more than three times  $p$.
Note that, in the case of $W$=0, the critical value of $J$ for
binding a hole of the  appropriate spin to a Mn ion is $J_c = 3.95677t$. 
For values of $J>J_c$, the holes form a parallel-spin impurity band split from 
the main general band \cite{Chatto}.
In that case, the physics becomes more similar to that of the Double-Exchange materials. 
In our model,  the parameter $W$ allows us to modulate the hybridization between
the Mn ions orbitals and the host semiconductor band. 
In the case of $J$=3$t$, we believe that a  lower impurity band is not formed  for 
any positive value of $W$. 

\begin{figure}
\includegraphics [clip,width=8.cm]{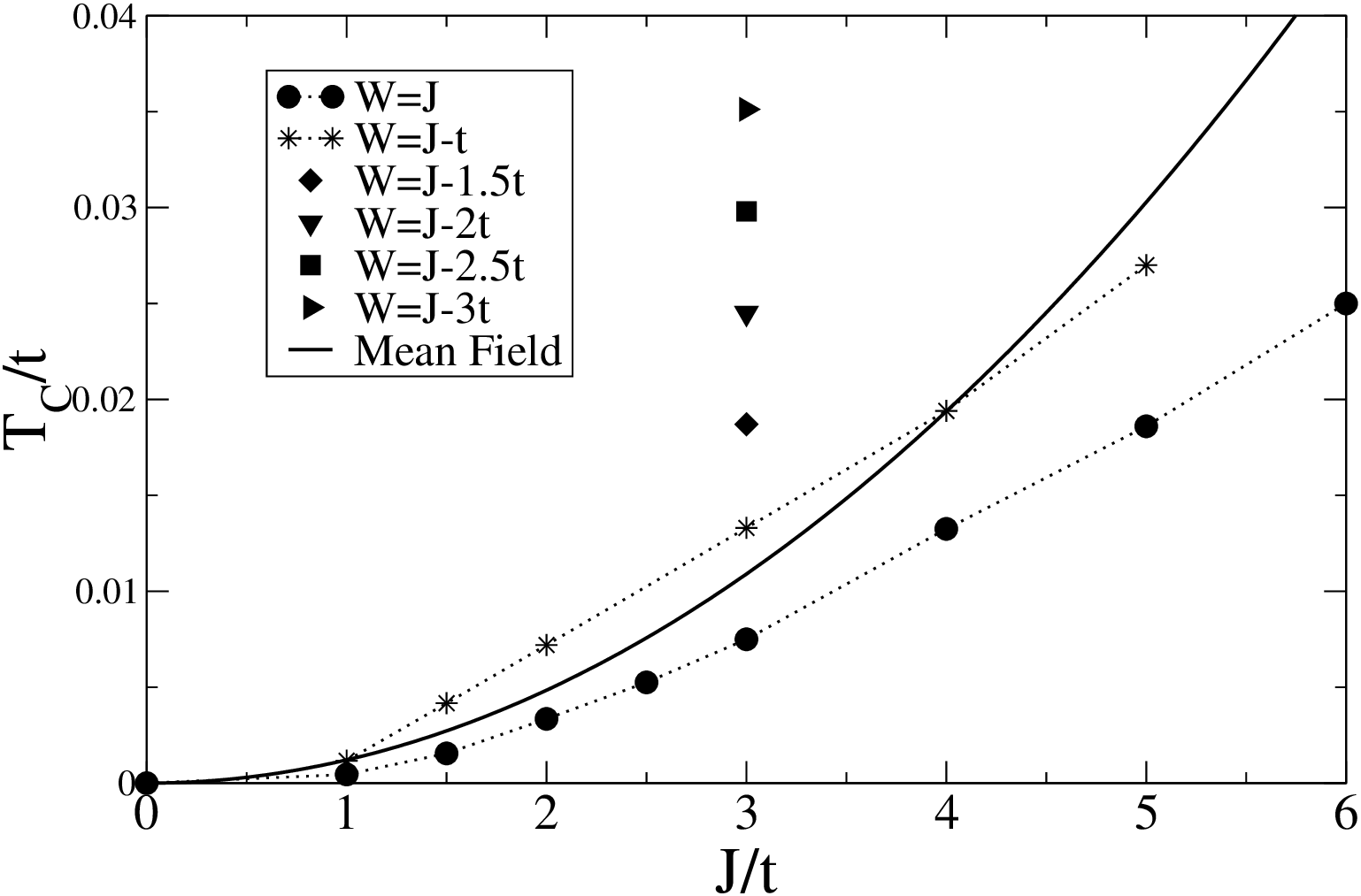}
\caption{Critical temperature as a function of $J$ for different values 
of $W$. 
The results corresponds to the 
case  $x$=10$\%$, $p$=0.7$\%$, and $N$=14.
The Weiss mean field  critical temperature is plotted.
} 
\label{fig4}
\end{figure}

\begin{figure}
\includegraphics [clip,width=8.cm]{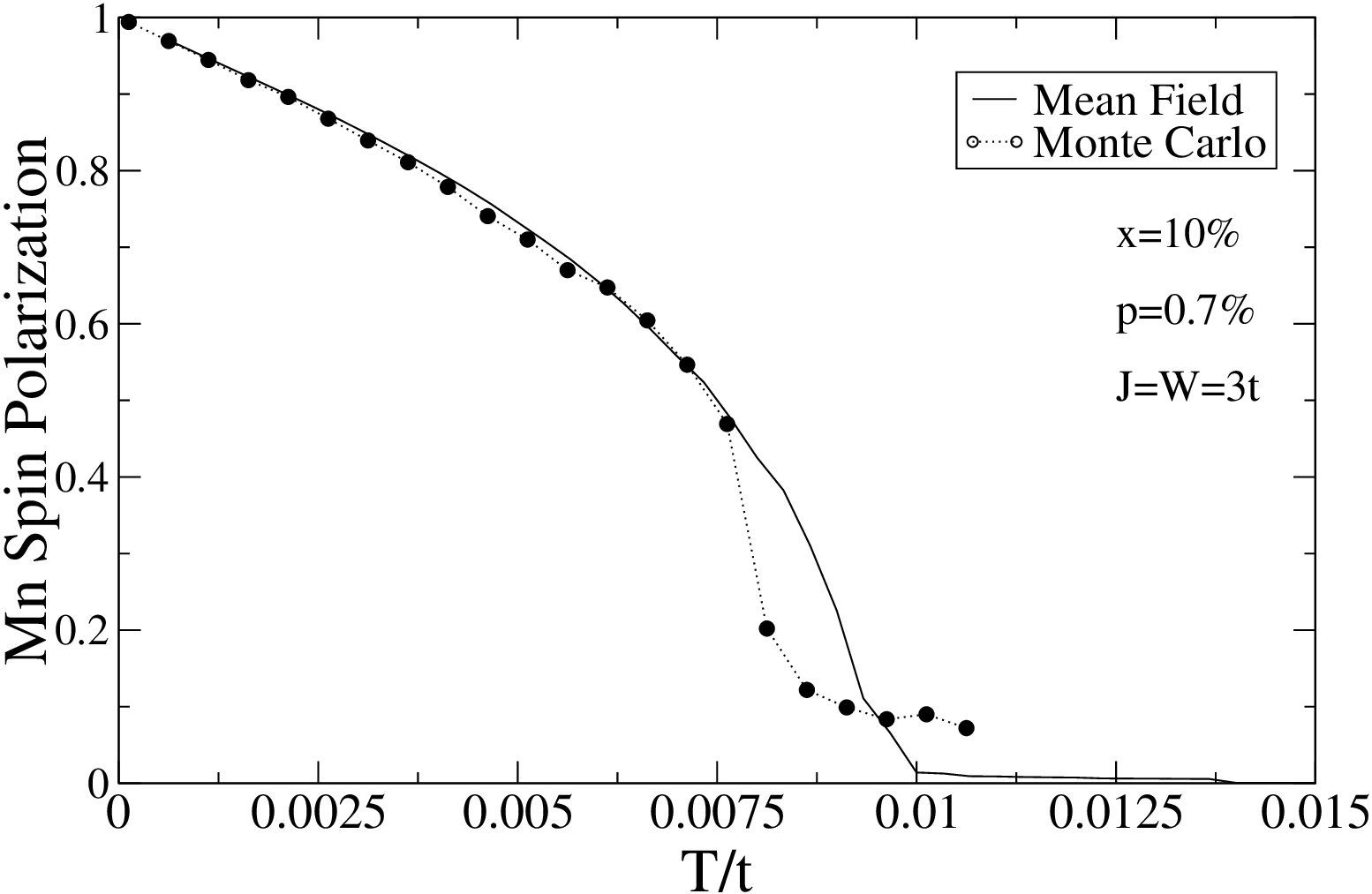}
\caption{Mn ions spin polarization as a function of $T$ 
as obtained from Monte Carlo simulations  and from Weiss mean field  theory.
The results correspond to $x=10\%$, $p=0.7\%$, $J$=$W$=3$t$. The MC results have
been obtained in a cell of size $N$=14. 
} 
\label{fig5}
\end{figure}

In Fig.~\ref{fig4} we plot the critical temperature obtained from MC simulations as
a function of $J$ for different values of $W\!$, and for concentrations $p$=0.7$\%$ and 
$x$=10$\%$. We have also plotted the critical temperature, Eq.~\ref{TcVCA},  obtained in the 
Weiss mean field
 theory \cite{Dietl,Dietl-bis}. 
The Weiss mean field $T_c$ increases quadratically with $J$ and, as explained before,
the approximation
treats the magnetic impurities in the virtual crystal approximation, leading to a
$T_c ^{VCA}$  independent of $W$. From the MC results, we observe that 
the MC critical temperatures are smaller than the $T_c ^{VCA}$'s only for 
values of $W$ very close to $J$. 
This is because, for $W$=$J$, the effective impurity potential is zero, and
the virtual crystal approximation  for obtaining
the Curie temperature seems to work well.  In this case, $W$=$J$ and small $J$, the thermal fluctuation correction
to 
$T_c ^{VCA}$ is similar to that occurring in the standard  Heisenberg model,
$\sim 25\%$.
This becomes clear in Fig.~\ref{fig5}, where we plot $M(T)$ as obtained in MC simulation and compare
it with the predictions of Weiss mean field. 
Both curves are very similar at low $T$, and only near the Curie temperature do they separate,
 due to thermal fluctuation effects.
For small values of $W\!$, the MC critical temperatures are larger than the $T_c ^{VCA}$'s.
This is due to the charge located on  the Mn ions which increases the
effective magnetic field felt by the Mn spins.

If the effective impurity potential $J$-$W$ increases,  holes would eventually form a 
localized  impurity band, leading to a decrease of  $T_c$. 
As commented above, we do not get close enough to this limit, whose 
MC description would require a large basis beyond our present computational capabilities.

\begin{figure}
\includegraphics [clip,width=8.cm]{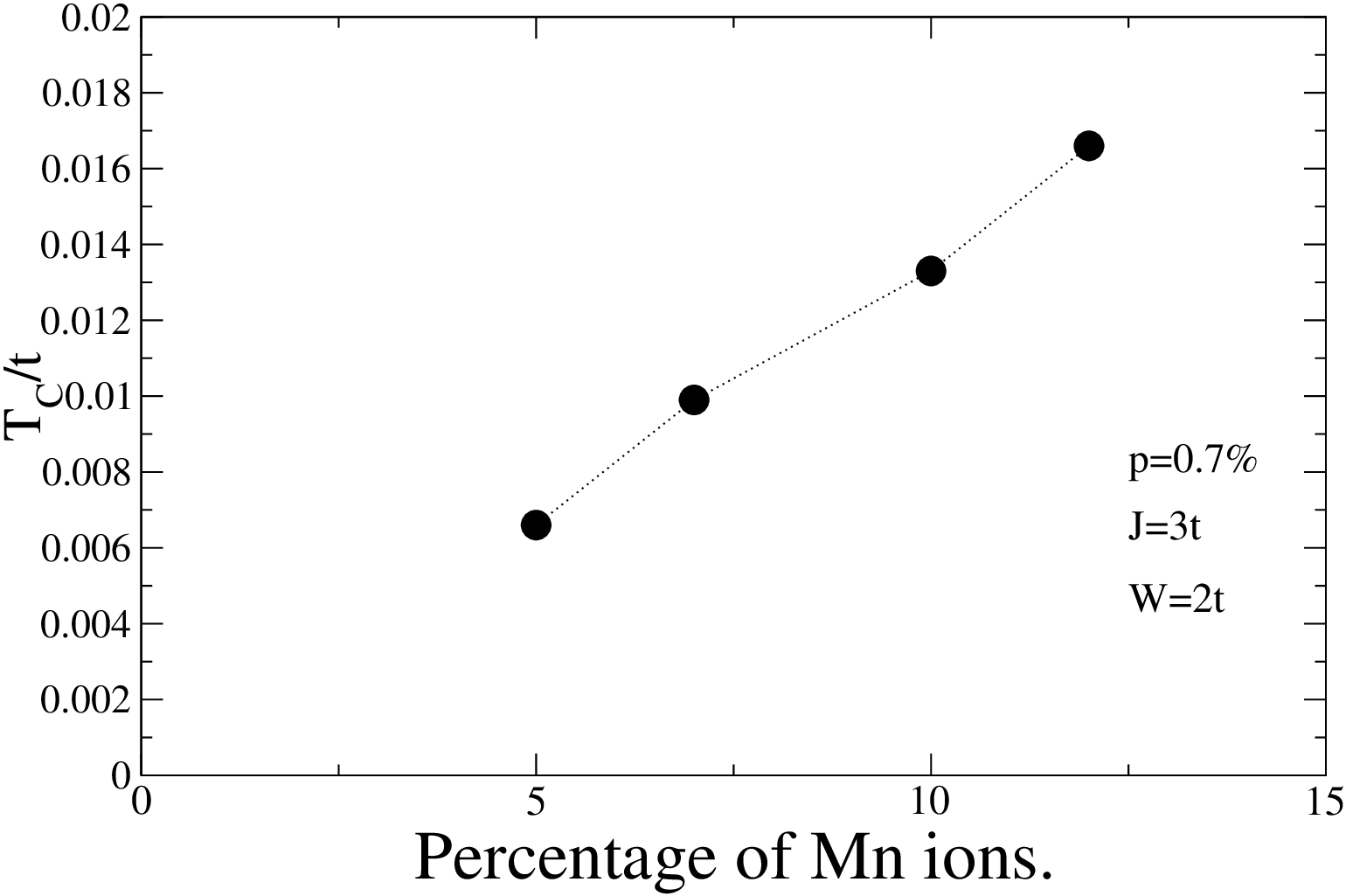}
\caption{Curie temperature as a function of the Mn ion's concentration
as obtained from Monte Carlo simulations. 
The results correspond to $p=0.7\%$,  $J$=3$t$ and $W$=2$t$. The MC simulations  have
been done  in a cell of size $N$=14. 
} 
\label{fig6}
\end{figure}

We have also studied the dependence of the critical temperature 
on the Mn concentration, $x$. This is shown in Fig.~\ref{fig6}, where we plot $T_c$ as a function of $x$ for
$J$=3$t$, $W$=2$t$, and $p$=0.7$\%$. Within the error margin, the
critical temperature increases linearly with $x$, as expected for low impurity concentration.

Finally, we have studied the dependence of $T_c$ on the hole charge density $p$. In Fig.~\ref{fig7}
we plot the MC results corresponding to the case $J$=3$t$, $x=10\%$, and two different
values of $W$. For comparison, we have also plotted the
Weiss mean field result. The Weiss mean field  predicts a $p^{1/3}$ 
dependence,  at low hole density. The MC results could be fitted perfectly to this dependence.
Again we obtain that the Weiss mean field $T_c$'s are higher that the MC results 
only for values of $W$ near the value of the exchange coupling $J$.

Before finishing our discussion on the MC results, a comparison with previous
simulations  is in order.
Recently, Schliemann and coworkers \cite{Schliemann}
have presented some MC results for (III,Mn)V semiconductors. They work in a model
similar to ours, but with the site energy difference set to zero, $W$=0. They describe the
motion of the carriers  using a plane wave basis with a cutoff $k_c$, 
and the exchange potential created by
the magnetic impurities is modeled with a Gaussian of thickness $a_0$. They obtain  that
the critical temperature is substantially smaller than the estimated using Weiss mean field theory.
This is  opposite to our results: for $W$=0 
our MC $T_c$ is higher than the mean field value.
As the exchange constant and the densities  (Mn  and carriers) used
in Ref.~\cite{Schliemann} are similar
to ours, there is a clear contradiction between our results and 
those of reference \cite{Schliemann}. A possible explanation for this discrepancy can be found in   
the wavelength  cutoff used in ref(\cite{Schliemann}), considerable 
larger  than the impurity Gaussian width, and, perhaps,  unable 
to trap carriers charge. If this were the case, their results could best correspond to 
our calculation for $W$=$J$, for which $T_c$ is smaller than $T_c^{VCA}$.   
In reference ({\cite{Schliemann}), the wavelength cutoff is near 25\AA, whereas
the value of $a_0$ is smaller than 5\AA. 

A comparison with experimental results is
 strongly dependent on the values of the parameters. For instance, 
 in the case of $t$=0.85$eV$, $J$=3$t$, $x$=10$\%$, 
and $p$=0.7$\%$, the critical temperature varies between 80K for $W$=$J$, and 350$K$ for $W$=0.
This range would agree rather well with
experimental data, assuming values of $W$  near $J$.
Given the oversimplified nature of our model, this comparison with experiments cannot be taken 
too far away. 
In any case, our results indicate clearly the relevant role of value of $W$
on the magnetic transition: decreasing its value, if possible experimentally,
would increase considerably the Curie temperature.
\begin{figure}
\includegraphics [clip,width=8.cm]{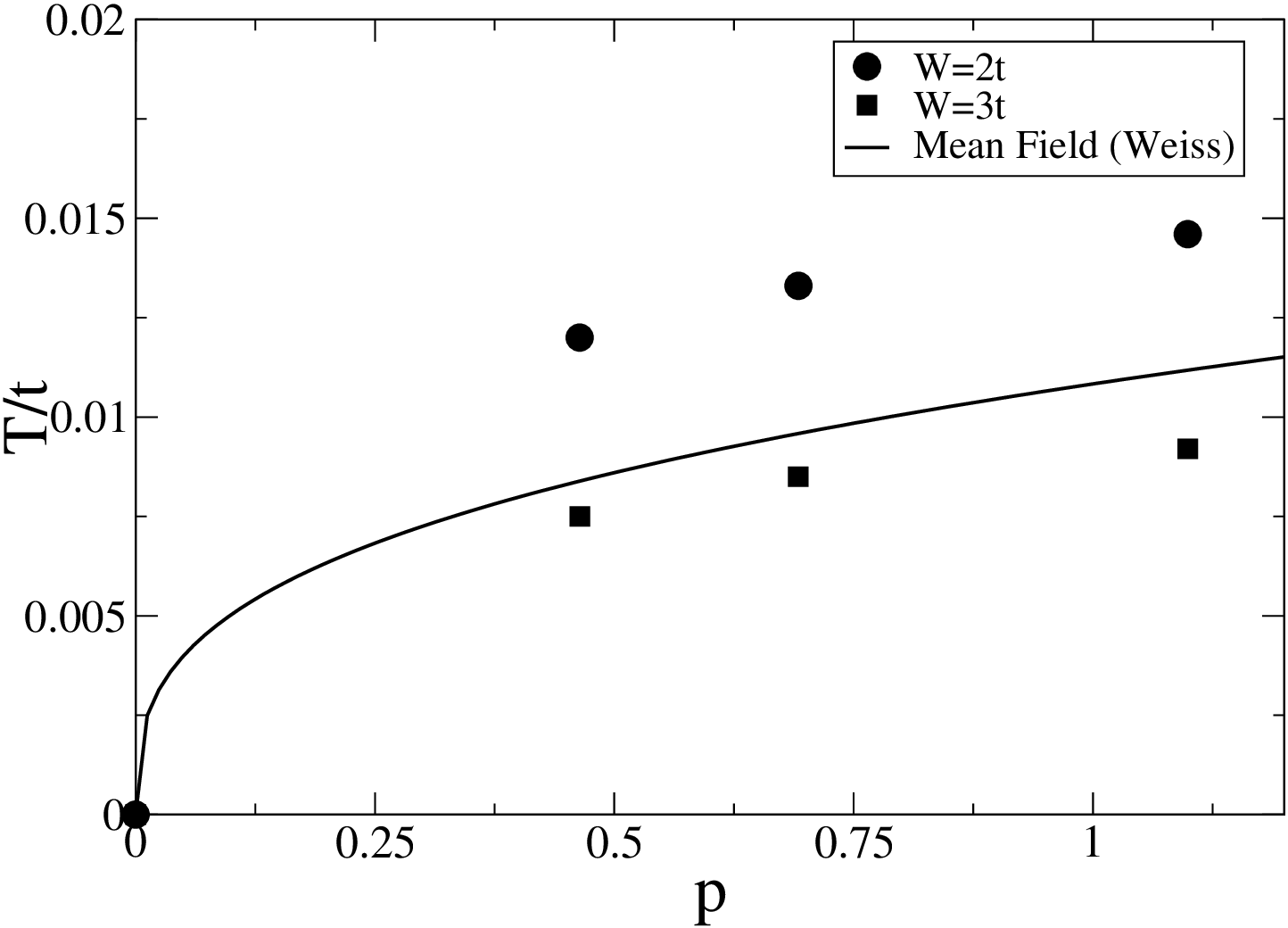}
\caption{Curie temperature as a function of the hole  concentration, $p$, 
as obtained from Monte Carlo simulations  and from Weiss mean field  theory.
The results correspond to $x=10\%$ and  $J$=3$t$. The MC simulations  have
been done  in cells of size $N$=12, $N$=14, and $N$=16. 
} 
\label{fig7}
\end{figure}

\subsection{CPA results.}

 In this section, we present results for the phase diagram obtained with the CPA
 method explained in \ref{CPA}. CPA results are parameterized by the
 density of states of the bare system,  chosen to be that of the
 {\em simple cubic} lattice. Although unnecessary from the CPA point of view,
 we have assumed degenerate fermions.  This is both physically correct for the
 relevant range of parameters, and makes  the comparison with MC results
 more meaningful. We have verified that the inclusion of fermion temperature
 does not change results in any significant way.
  
\begin{figure}
\includegraphics [clip,width=8.cm]{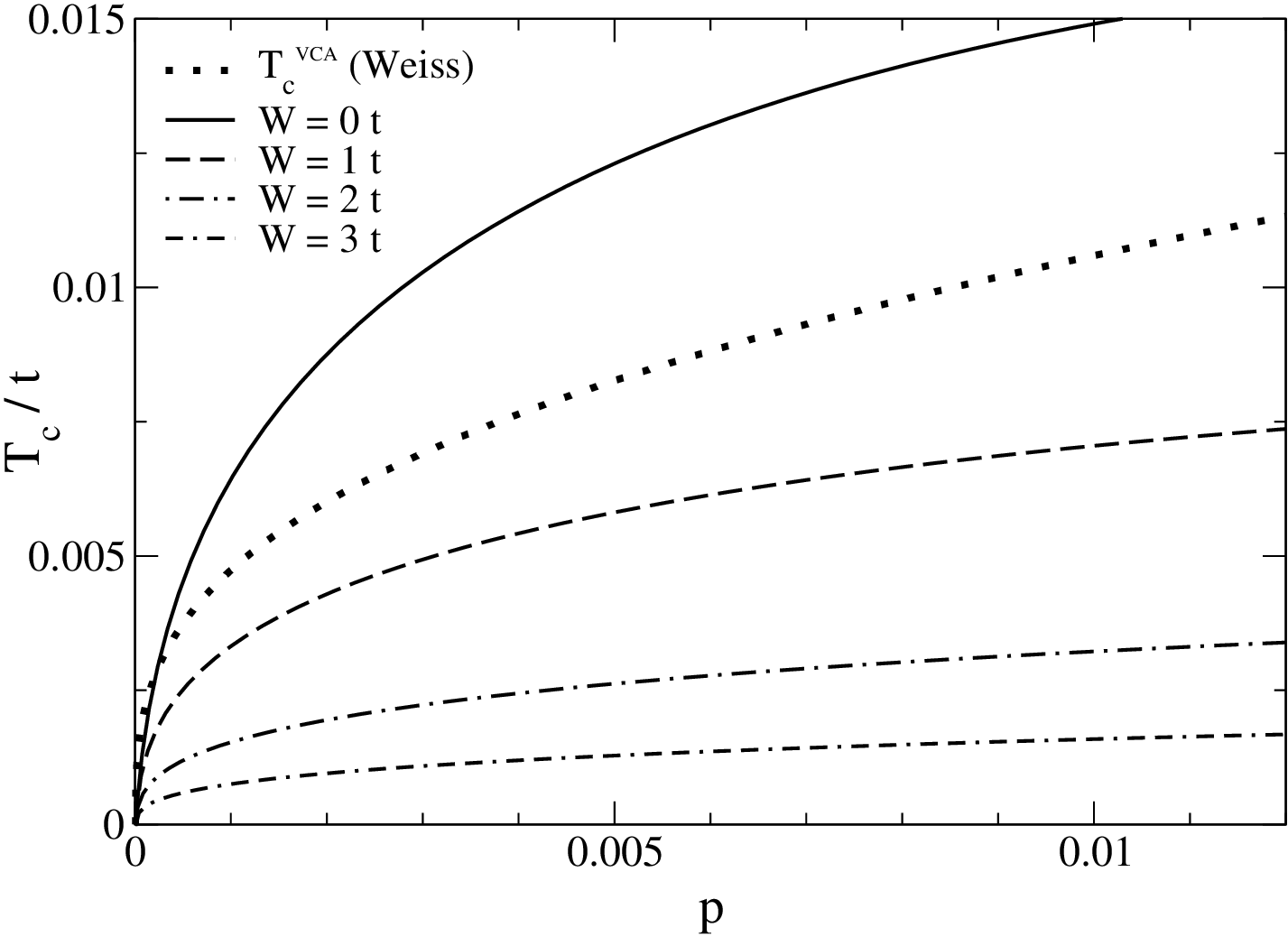}
\caption{\label{fig:cpa1} CPA critical temperature
 as a function of hole concentration for $x=0.1$,  
 $J=3t$, and different  values of $W$. The Weiss (VCA) mean field critical temperature
 is also plotted for comparison.
}
\end{figure}

 In fig.~\ref{fig:cpa1} we present the critical temperature for $J=3t$ as a function of carrier
 concentration in the relevant region, for several values of $W$. 
 As in the  MC calculations  (see Fig.~\ref{fig4}),
 the most important effect of $W$ is the depression of $T_c$.
 Notice, though, that the overall temperature scale of the
 CPA calculation is {\it lower} than that of the MC results (compare with Fig.~\ref{fig7}).
  This
 reverses the trend found in most statistical mechanics models, where the MC
 (exact) transition temperatures are  higher that the mean field
 counterparts.  We ignore whether this effect is a genuine peculiarity of our
 model or rather the outcome of  approximations. We observe that  $T_c$ follows 
 the $p^{1/3}$
 dependence with carrier concentration expected to hold close to a 3-d band  
 in the Weiss mean field  
 approximation, but  this agreement between CPA and VCA is restricted to the small
 carrier concentration region (see below). In addition, the very apparent effect of $W$
 on $T_c$ is completely missed in the VCA, as mentioned before.  
 
\begin{figure}
\includegraphics [clip,width=8.cm]{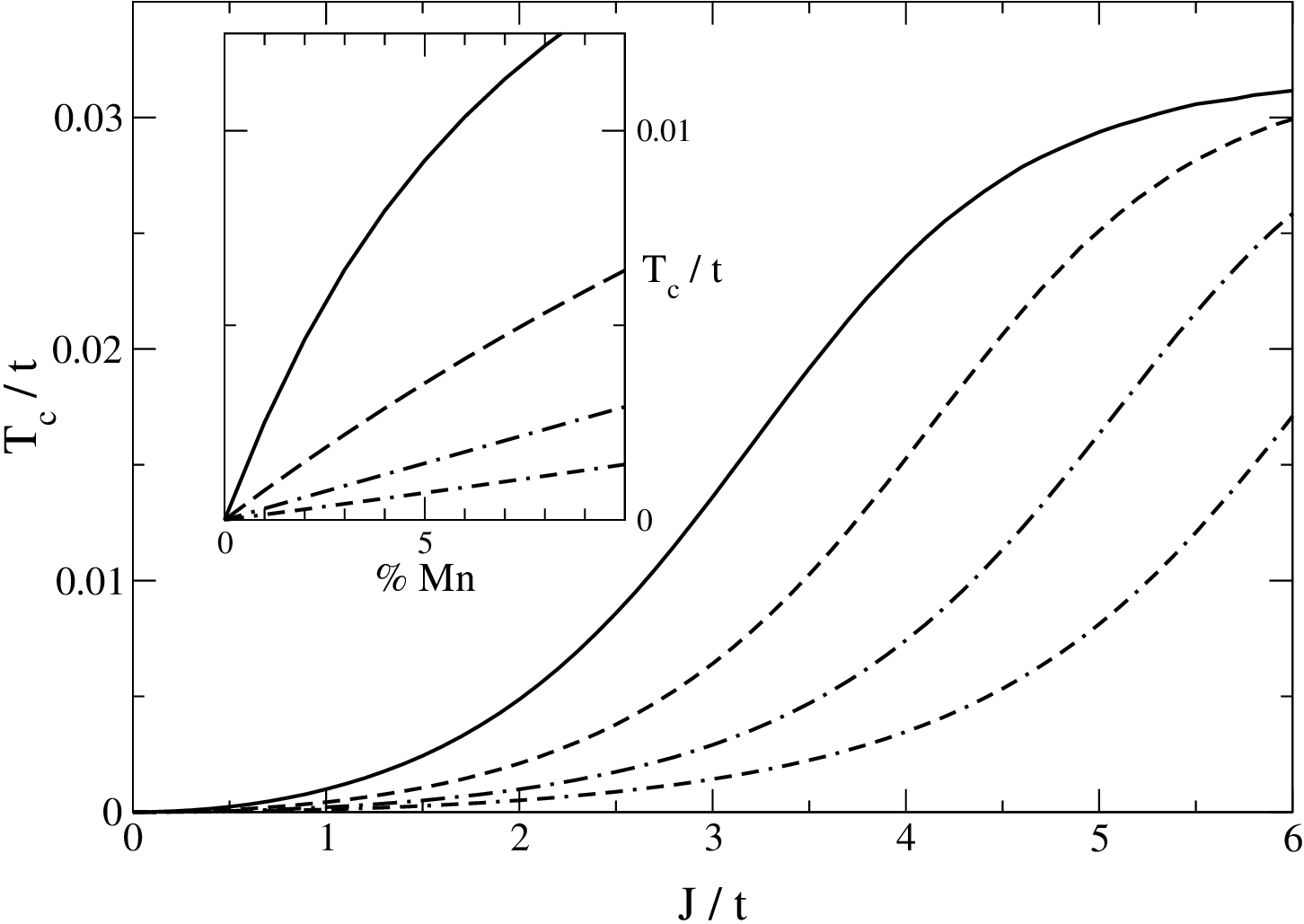}
\caption{ CPA critical temperature as a function of $J$ for $x=0.1$ and 
$p=0.007$. The different curves correspond to the following values of $W$
(top to bottom): $W=0t$, $1t$, $2t$, and $3t$.
Inset: Critical temperature as a function of impurity concentration  
for $J=3t$, $p=0.007$, and the same sequence of values of $W$ (top to bottom).}
\label{fig:cpa2}
\end{figure}

 $T_c$ as a function of Hund coupling $J$ is presented in fig.~\ref{fig:cpa2}, 
 for several values of
 $W$. The prominent effect of $W$ in reducing the critical temperature is very
 clear. Comparing with the MC data of Fig.~\ref{fig4}, we see  qualitatively 
 similar curves,    
 but, again,  the MC temperatures are higher than the CPA ones.
 It is worth comparing the CPA results of fig.~\ref{fig:cpa2} with those of the
  VCA (Eq.~\ref{TcVCA}), which predicts a
 dependence of the form: $ T_c^{VCA} \propto x J^2$, irrespective of the size of $J$ and $x$.
 In the CPA treatment, this
 dependency only holds for small values of both  $J$ and $x$, 
  and significant deviations from the $J^2$ law are already evident in fig.~\ref{fig:cpa2},
  particularly for $W=0$.
 We have also studied the dependence of $T_c$ with
 impurity density, as shown in the inset of fig.~\ref{fig:cpa2} for small values of $x$.
 As in the MC calculations, an approximately linear regime is obtained in this limit. 
  This linearity 
 suggests that the effective magnetic interaction between core-spins
could be well approximated by a two-body interaction in this regime.
  Notice that, in spite of the VCA prediction 
 (Eq.~\ref{TcVCA}), 
 this linear dependence  does not hold
for higher impurity density, or even in the concentration range of fig.~\ref{fig:cpa2}
 for $W=0$, where non linear effects are already explicit. 
 This behavior with $J$ and $x$ has been observed before for the present problem 
 in a CPA calculation with $W=0$ and a model semicircular density of states \cite{Chatto}.
  Its survival
 upon inclusion of disorder $W \neq 0$ and for the {\it real} density of states (sc lattice)
 confirms that it is a robust feature of the model.

\begin{figure}
\includegraphics [width=8.cm]{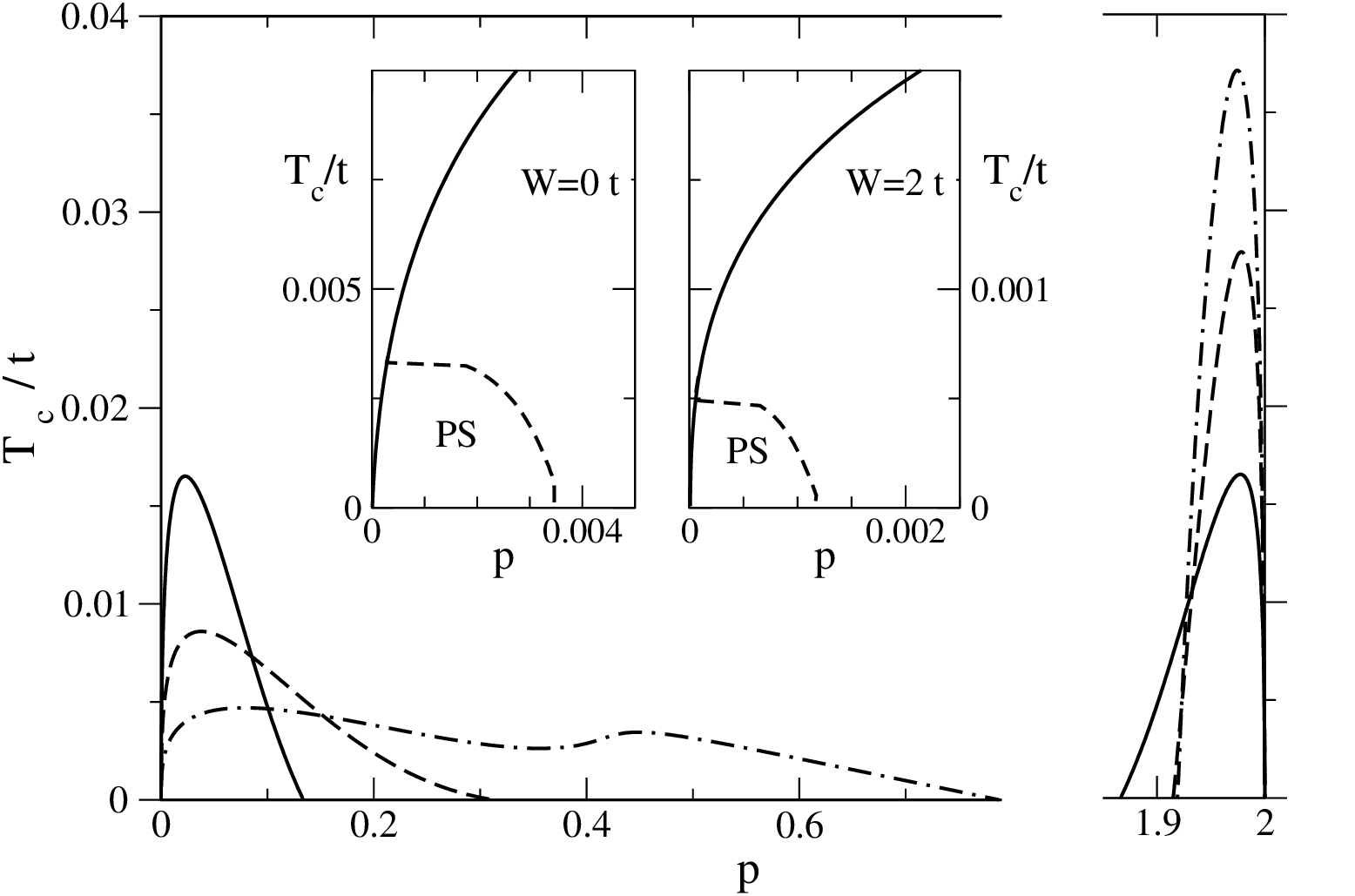}
\caption{ CPA critical temperature as a function of hole density for 
 $J=3t$ and the following values of $W$: 
 $W=0t$ (continuous line), $W=1t$ (dashed line), and $W=2t$ (dashed-dotted line).
Insets: Enlarged view of the phase diagrams showing the regions
  of phase separation for $J=3t$, with $W=0t$ (left inset) and $W=2t$
  (right inset).}
\label{fig:cpa3}
\end{figure}

 It has been argued earlier that the effect of $W$ on $T_c$ can be  correlated
 with  changes in carrier density on the impurity sites. Being the magnetic coupling
 mediated by mobile carriers, decreasing their density on the impurity sites (with
 the repulsive term $W$) should lead to a decrease in $T_c$. We can test  
  this behavior analyzing the  limit of an almost filled band,
 with electrons as carriers (remember that our bare fermions are holes).
  The effect of $W$ would then be the opposite and, therefore, $T_c$ should
  increase. To verify this idea while gaining a better perspective of the
  problem, we have plotted in fig.~\ref{fig:cpa3} the phase diagram 
  in the whole range of carrier concentration, for three values of $W$.
   For $W=0$, the problem ($T_c$) is
  particle-hole symmetric, and the ferromagnetic instability is restricted to the band
  edges. 
  Although quantitatively far from the small $J$ limit, this confinement to band
  edges is qualitatively similar to
 the RKKY
 regime, where the same behavior can be  traced back to the oscillatory
  nature of the magnetic coupling.  
    As anticipated,  fig.~\ref{fig:cpa3} confirms the  asymmetric effect   
   of $W$ on both types of carriers: it decreases $T_c$ at the
   lower edge (holes) and increases $T_c$ at the upper edge (electrons). 
   Notice that, for electrons, $W$ acts as an attractive impurity, increasing the
    carrier density in the impurity sites, therefore increasing $T_c$.
  It is worth mentioning that fig.~\ref{fig:cpa3} also contains the CPA results for 
  $W < 0$:   hole density $ p $  and $ W > 0 $ is equivalent to 
  having chosen  $ W < 0$  and hole density $2 - p$ ({\it i.e.} electron density $p$). From
  this point of view, the physics close to the upper  end of fig.~\ref{fig:cpa3} upon
  increasing the value of $W$ is controlled by the development of a magnetic impurity band, 
  with magnetic behavior becoming closer to that of the Double-Exchange model \cite{Chatto}. 
   Another interesting effect of increasing $W$ shows up in  fig.~\ref{fig:cpa3}: 
   the spread of the magnetic instability 
  from the lower  edge into the band. We cannot  offer a simple  
  explanation for this fact.
%  We tentatively associate this behavior to the disorder
%  effect of $W$ (effective both for electrons and holes) which would wash out 
%  Fermi surface effect, that is, the oscillatory nature of magnetic coupling. 

   Finally, we have investigated the phase diagram looking for the presence of phase 
   separation (PS). PS is a common feature
   in magnetic models with itinerant carriers \cite{n97,ag98,detal98,
   agg99,myd99}. It seems to 
   appear close to the magnetic transition temperature, when $T_c$ changes
   abruptly with carrier concentration \cite{Guinea}.
    In order to search for PS, we explore the existence
   of {\it global minima} in the thermodynamic potential, in addition to the usual 
   {\it local minimum}
   employed to obtain the curve of $T_c$.
    We observe small regions,  close to the band edges,where the systems is unstable
   versus PS. This is shown  in the insets of fig.~\ref{fig:cpa3} for two values of $W$.
    Comparing results for $W=0$ and $W=2t$, we see that 
   the PS region decreases with increasing
   $W$, due to the 
   associated reduction in the dependence of $T_c$
   with carrier concentration. In the limit of small electron 
   concentration, similar PS areas (not shown here) 
   are found with opposite behavior with  $W$.

   As a final remark, we note that our calculation becomes qualitatively 
   similar to  a recent CPA analysis \cite{Chatto}, carried out in the absence of
    impurity potential: $W=0$.  Quantitatively, however, it is
   different (even for $W=0$) because of our using of the real density of states
   (sc lattice) as opposed to a model semicircular one. Nevertheless, close to
   the band edges, the differences between both calculations for $W=0$ amount 
   to a different choice of  lattice regularization, or, equivalently,
    a rescaling of the   Hund coupling $J$.

\section{Summary.}

We have studied a simple model for diluted magnetic semiconductors. In this 
model the carriers motion is described by a tight binding Hamiltonian and the
hole-Mn ion coupling  has two contributions: an antiferromagnetic exchange interaction 
between the Mn spin and the carrier spin, $J$, and an electrostatic interaction, $W$,
between the carriers charge and the potential arising from the Mn dopants.
We have studied this model performing Monte Carlo simulations and mean field
CPA calculations.  
Our main results are the following:
\par \noindent 1) The hole charge on the Mn ions
has a relevant effect on the Curie temperature. The effective impurity potential $J-W$ 
controls
the hybridization of the Mn impurities with the host bands and the hole charge
on the impurities. By increasing $J-W$, if possible experimentally, 
the charge on the Mn ions
and the critical temperature would increase considerably.
\par \noindent 2) 
The Monte Carlo critical temperature is always higher than the corresponding CPA
temperature. We also find that the Monte Carlo critical temperature is in general higher than the 
obtained from the simplest Weiss mean field theory. Only for values of $J-W$ near cero
the Weiss $T_c$ is higher than the Monte Carlo one.
\par \noindent 3) The CPA calculation shows the existence of small pockets
of phase segregation instability close to band edges.

\section{Acknowledgments.}
We thank Prof. F. Guinea and Dr. J. Fern\'andez Rossier for discussions. Finantial
support is acknowledged from Grants No PB96-0085 (MEC, Spain) and CAM-07N/0008/2001 
and CAM-07N/0015/2001  
(Madrid, Spain)

%%%%%%%%%%%%%%%%%%%%%%%%%%%%%%%%%%%%%%%%%%%%%%%

%\bibliography{mia}

\begin{thebibliography}{53}
\bibitem[{not()}]{nota1}
\bibinfo{note}{In the interaction between the ion spins and the fictitious
  field ${\mathbf B}_ f (I)$ we take $g \mu _B =1 $}.
\end{thebibliography}

\end{document}